# Application Component Placement in NFV-based Hybrid Cloud/Fog Systems with Mobile Fog Nodes[+]

Carla Mouradian, Somayeh Kianpisheh, Mohammad Abu-Lebdeh, Fereshteh Ebrahimnezhad, Narjes Tahghigh Jahromi, Roch H. Glitho

CIISE, Concordia University, Montreal, QC, Canada

{ca_moura, s_kianpi, m_abuleb, f_ebrah, n_tahghi, glitho}@encs.concordia.ca

*Abstract*— **Fog computing reduces the latency induced by distant clouds by enabling the deployment of some application components at the edge of the network, on fog nodes, while keeping others in the cloud. Application components can be implemented as Virtual Network Functions (VNFs) and their execution sequences can be modeled by a combination of sub-structures like sequence, parallel, selection, and loops. Efficient placement algorithms are required to map the application components onto the infrastructure nodes. Current solutions do not consider the mobility of fog nodes, a phenomenon which may happen in real systems. In this paper, we use the random waypoint mobility model for fog nodes to calculate the expected makespan and application execution cost. We then model the problem as an Integer Linear Programming (ILP) formulation which minimizes an aggregated weighted function of the makespan and cost. We propose a Tabu Search-based Component Placement (TSCP) algorithm to find sub-optimal placements. The results show that the proposed algorithm improves the makespan and the application execution cost.**

1. INTRODUCTION

Cloud computing comes with several inherent capabilities such as scalability, on-demand resource allocation, flexible pricing models, and easy application and services provisioning. However, the fundamental

---

limitation of cloud computing is the physical distance between a cloud service provider's data centers (e.g., Amazon Web Services (AWS), Google, etc.) and end devices. This distance could cause end-to-end delays which may not be acceptable for latency-sensitive applications. Well-known examples include disaster management, healthcare, and autonomous driving applications.

Fog computing [1] is a computing paradigm introduced to tackle the cloud latency-related challenge. Indeed, it extends the traditional cloud computing architecture to the edge of the network, enabling computing at the edge of the network, closer to IoT and/or the end-user devices.

Extending cloud computing to the edge of the network results in a hybrid cloud/fog system. Such hybrid system allows the processing of some application components (e.g., latency-sensitive ones) at the edge of the network, by the so-called fog nodes, while processing the others (e.g., delay-tolerant and computationally intensive components) in the cloud. Fog nodes can be either static or mobile. For instance, a drone can act as a mobile fog node [2]. A more general term of nomadic data centers has been introduced in [3]. Nomadic data centers can also act as mobile fog nodes. They denote small, portable edge data centers that can collect and process data.

Applications in hybrid cloud/fog systems can be implemented as a set of interacting components. These components together form structured graphs with the following sub-structures: sequence, parallel, selection, and loop [4]. The problem of application component placement over hybrid cloud/fog systems has been studied by several researchers. Most of these works have focused on application latency minimization (e.g., [5][6]). However, the main drawback of these works is that they assume that the fog nodes are static nodes with predefined locations. On the other hand, in real systems, a fog node can be mobile. Such level of mobility introduces new challenges to the component placement problem in hybrid cloud/fog systems. In order to place the application components on such mobile fog nodes, the stationary distribution for the locations of mobile fog nodes could be used to obtain their location at any time.

This paper focuses on application component placement in Network Function Virtualization (NFV) -based hybrid cloud/fog systems with mobile fog nodes, where the cloud and the fog infrastructures are provided as NFV Infrastructure (NFVI). We aim at the minimization of the aggregated weighted functions of applications makespan and cost (a budget for resource consumption). NFV is an emerging technology that employs virtualization as a key technology. It aims at decoupling the network functions from the underlying proprietary hardware and running them as software instances on general purpose hardware [7][8]. Application components in NFV-based hybrid cloud/fog systems can be implemented as Virtual Network Function (VNFs), e.g., [9][10][11]. The structured graphs representing the applications are therefore VNF Forwarding Graphs (VNF-FG) (i.e., sets of VNFs chained in specific orders). Fig. 1 illustrates an example of earthquake early warning and recovery application. It shows the components of the application and their structured VNF-FG representation with sequence, selection, and parallel sub-structures.

In NFV settings, the application component placement problem could be tackled as a VNF-FG embedding problem because it consists of mapping the structured VNF-FGs onto the NFVI. It should be noted that this paper assumes that a standard ETSI NFV framework is used. In such a framework, it is the NFV Management and Orchestration (MANO) functional entity that orchestrates and manages the VNFs [12]. The algorithm proposed in this paper is a potential algorithm that could handle the placement problem in an ETSI MANO since ETSI considers the algorithmic issues as implementation issues and does not standardize them. There are several open sources for implementing MANO functional entity and we consider the implementation of a full-fledged MANO that runs our proposed algorithm outside the scope of this paper. However, we have indeed validated the proposed algorithm. It should also be noted that our proposed solution will fit very well in any other orchestration framework in which component placement algorithms are needed.

The VNF-FG embedding problem has received significant attention in the research community [13]. Various objectives have been considered in the literature, such as operational cost [14], number of VNF instances [15], and resource utilization [16]. To the best of our knowledge, the only VNF-FG embedding solution that considers hybrid cloud/fog systems and is applicable to structured VNF-FGs with various sub-structures such as selection and loop is our previous work [4]. In fact, this paper is an extension of our previous work [4]. However, in [4] we assumed all fog nodes are static, an assumption which may not hold in real scenarios. In contrast, in this paper, we consider scenarios where the fog nodes can be mobile and mathematically model this mobility using the Random Waypoint (RWP) model [17]. In addition, our previous work [4] lacks a heuristic. It advocates an exhaustive search to solve the placement optimization problem. This is time-consuming and not scalable for large applications and/or a large number of cloud/fog nodes. In contrast, this paper proposes a Tabu Search-based meta-heuristic to find a sub-optimal solution for large-scale scenarios in feasible time. The main contributions of the paper can be summarized as follows:

- Modeling fog mobility with the RWP model and calculating component execution time and cost; aggregating the calculations to obtain the application makespan and cost.
- Formulating the application component placement problem as an Integer Linear Programming (ILP) problem which minimizes the weighted aggregated function of makespan and cost.
- Proposing Tabu Search-based Component Placement (TSCP) algorithm to find sub-optimal solution.

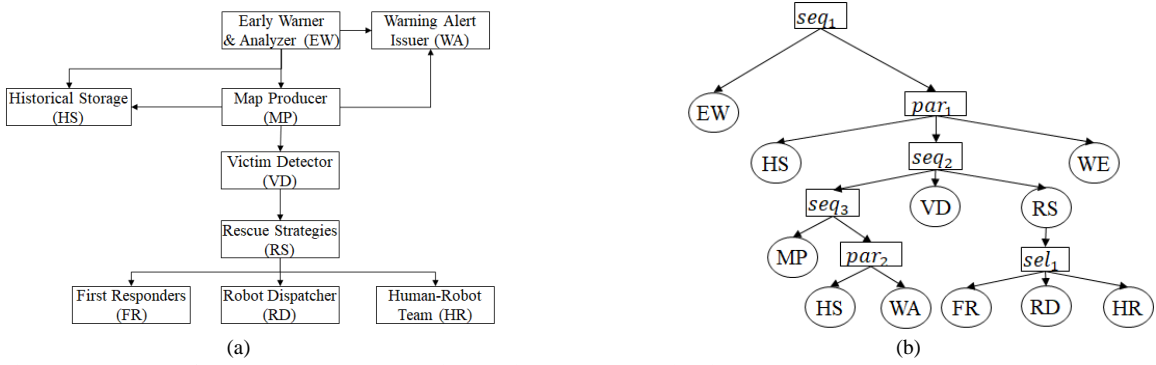

Fig. 1. Example of earthquake early warning and recovery application
(a) Component-based application
(b) Structured VNF-FG representation

The rest of the paper is organized as follows: Section 2 presents the motivating scenario and reviews the literature. The system model is explained in Section 3. Section 4 presents the mobility model and the optimization problem. The Tabu Search-based algorithm is described in Section 5, followed by the evaluation results in Section 6. We conclude the paper and outline likely future work in Section 7.

## 2. MOTIVATING SCENARIO AND LITERATURE REVIEW

### 2.1. Motivating Scenario

An earthquake application can illustrate the motivation behind our work. One example is the large-scale earthquake that hit Kobe, Japan, in 1995. Measuring 6.9 magnitudes. It left more than 5000 deaths and 13,000 injuries. The application performs rapid identification of disasters and monitoring of disaster-prone areas. The monitoring can be facilitated by UAVs such as drones. Drones are usually equipped with camera sensors that can capture images. The collected images are processed later to produce damage-assessment maps to allow responders to serve areas that experience damage to critical areas first.

This application can be composed of several components, i.e., Fig. 1. For instance, *Early Warner and Analyzer (EW)* processes the data received by different sensors such as seismic and camera sensors, and accordingly detect prospective danger. The selected data are sent to *Historical Storage (HS)* for long-term storage and analysis and to *Warning Alert Issuer (WA)* for public warnings. In addition, a *Map Producer (MP)* processes these to find the epicenter location and produce damage-assessment maps. These maps are also used by the *Victim Detector (VD)* to locate possible human being presence. When victims are detected, the *Rescue Strategies (RS)* is informed to take immediate life-saving decisions. The *RS* instructs either *First Responders (FR), Robot Dispatcher (RD),* or *Human-Robot Team (HR)* to begin the rescue missions.

We consider a system with three layers: an IoT/end-users layer that can include different types of IoT devices, such as seismic sensors; a fog layer that can include both mobile fog nodes such as drones and static fog nodes, e.g., relief vehicles; and a cloud layer consisting of the distant data centers. Accordingly, some of the application components, specifically the latency sensitive ones, could run on the fog layer, with some running on mobile fog nodes such as the *Victim Detector*, while others on static fog nodes, such as *Warning*

*Alert Issuer*. Computationally-intensive and delay-tolerant components could run in the cloud layer; these might include the *Historical Storage* and *Map Producer* components.

## 2.2. Literature Review

In this subsection, we review the relevant literature on application component placement over hybrid cloud/fog NFVIs. In the first subsection, we review the proposed solutions for application component placement in hybrid cloud/fog systems where these components are not placed as VNFs. We then review the works to date on VNF-FG embedding that do not focus on hybrid cloud/fog systems. To the best of our knowledge, our previous work (i.e., [4]) is the only one that investigates the placement of application components as VNFs in NFV-based hybrid cloud/fog systems.

### 2.2.1. Application Component Placement in Hybrid Cloud/Fog Systems

Most of the proposed solutions for component placement over hybrid cloud/fog systems consider static fog nodes, such as access points, road side units. Few works have considered mobile fog nodes, such as a vehicle. In this subsection, we first review the proposed solutions that consider static fog nodes, and then we describe the proposed mechanisms that consider the mobility of the fog nodes.

#### a. Application Component Placement Considering Static Fog Nodes

Different objectives have been considered in the literature for application component placement over cloud/fog infrastructures. Mahmud *et al.* [18] consider placing components over cloud and fog nodes such that the user's Quality of Experience (QoE) is maximized. In contrast, Deng *et al.* [19] do so with the objective of minimizing the power consumption of cloud and fog nodes while considering the delay at the user's side. Many authors seek to minimize the application response time. Yin *et al.* [20] schedule the tasks over the cloud/fog infrastructures with the objective of reducing the response time of the tasks under a specified threshold. Similarly, Pham *et al.* [21] schedule the tasks over the cloud/fog system. They aim at minimizing the monetary cost of the rented cloud resources and the execution time of a workflow consisting of several interacting tasks.

In contrast to Pham *et al.*, where the application tasks interact through sequential and parallel sub-structures, in [4], we consider non-deterministic applications with more complex sub-structures such as selection and loop.

Other objectives have been considered besides optimizing the response time. Agarwal *et al.* [5] propose an algorithm that distributes the workload over the hybrid cloud/fog system while considering the throughput maximization and the response time minimization. Taneja *et al.* [6] do so such that the resources are utilized in an efficient manner and the application response time is minimized. In [22], Skarlat *et al.* aim at optimizing the utilization of fog nodes while satisfying the application QoS in terms of execution time. Authors in [23] and [24] tackle the problem from the perspectives of mobile devices. Hassan *et al.* [23] consider offloading of application tasks from mobile devices to cloud and fog nodes with the goal of minimizing the application execution time. Similarly, Bittencourt *et al.* in [24] schedule the workload offloaded by mobile users over cloud and fog infrastructures. They present different scheduling strategies to cope with applications with different objectives, such as a delay-priority strategy that prioritizes latency-sensitive applications.

Although these solutions address the problem of application component placement in hybrid cloud/fog systems, they only consider static fog nodes. This assumption makes their approach nonfunctional when the system includes mobile fog nodes such as vehicles, UAVs, or personal cell-phone devices.

  *b. Application Component Placement Considering Mobile Fog Nodes*

Very few works have considered the mobility of fog nodes when placing application components over the cloud/fog system. Zhu *et al.* [25] propose an algorithm to dynamically distribute the application tasks across static fog nodes (e.g., roadside units), mobile fog nodes (e.g., busses), and the cloud. Their goal is to find a balance between the application latency and quality loss. The proposed algorithm places individual tasks on cloud/fog nodes; however, in many real-world applications, there are interactions among an application's tasks that require an appropriate component placement mechanism. In our work, we place a set of interacting application tasks that can interact using multiple sub-structures (e.g., selection, loop).

Authors in [26] propose a computation offloading mechanism for mobile devices by using reinforcement learning. The tasks are offloaded to mobile fog nodes and to cloud nodes such that the service response time and the energy consumption of mobile devices are minimized. The proposed mechanism handles the mobility of the fog nodes by migrating a task from one mobile fog node to another whenever needed; however, it does not perform mobility aware offloading. In contrast, in this work, we use the probability distribution of the locations of mobile fog nodes to obtain their location distribution during placement decision.

*2.2.2. VNF-FGs embedding*

The problem of VNF-FGs embedding in NFV and cloud networks has been studied widely over the last few years. Various objectives have been considered, such as efficient infrastructure utilization [27][28], operational cost minimization [4][29], and provider revenue maximization [16][30]. In the following, we explain the solution approaches these works have used.

Moens *et al.* in [27] model the problem of placing a batch of VNF-FGs using ILP, which minimizes the infrastructure utilization. Fang *et al.* [28] propose a heuristic to solve the VNF-FG placement problem. They consider a balanced utilization of the spectrum of fiber links and infrastructure resources. Ghaznavi *et al.* [29] solve the problem with the objective of minimizing the operational cost of VNF placement while maintaining the QoS. In [4], we place application components as VNFs over hybrid cloud/fog NFVIs. However, in contrast to [29], we aim at minimizing the expected application makespan as well as the cost.

Authors in [30] and [16] maximize the provider's revenue. Sun *et al.* in [30] solve the problem by proposing online and offline methods. In the offline method, all requests are known in advance. The online method uses a prediction of future VNFs. Mechtri *et al.* [16] model the VNF-FG embedding problem as a weighted graph matching problem and propose an eigen-decomposition-based approach to solve it. These works are not directly applicable when fog resources are also involved since they implicitly assume that all resources are provided by the cloud. Indeed, using the fog brings two challenges to the problem. 1) The fog nodes' existence in the problem introduces a new type of heterogeneity compared to cloud resources; they have limited resources but provide faster response time. An appropriate allocation mechanism is required to exploit such resources. 2) Similar to what has been discussed about component placement approaches, the fog nodes can be mobile [31]. An efficient placement should consider such mobility to avoid high cost or a long makespan as a result of assuming static nodes with predefined locations. We propose a method that places application components containing various sub-structures such as sequence, parallel, selection, and loop in a hybrid cloud/fog NFVI where fog nodes can be mobile.

## 3. SYSTEM MODEL

In this section, we explain the modeling of the components implemented as VNFs, the structured VNF-FGs, the network, and the IoT/end-users that may interact with components.

**VNFs –** Each component of the application is implemented as a VNF. Let $T$ be the set of VNF types in the system. We denote the type $t$ of a VNF with $f^t$, which can be shared by more than one application. The resource requirements for processing a VNF $f^t$ per unit of resource (CPU, memory, storage) is represented by $\vartheta_{f^t}$ and the processing capacity of $f^t$ is represented by $c_{f^t}$. The set of available instances for $f^t$ is represented by $I_{f^t}$. Each VNF type $t$ has a predefined license cost $\partial_{f^t}$. We denote the maximum allowed VNF processing utilization with $\mu_{f^t}$.

**VNF-FG Requests -** Let $Req$ be the set of structured VNF-FG requests received by the system. We represent a single request with $R \in Req$. The set of required VNF types for request $R$ is indicated by $vnf_R$ ($vnf_R \subset T$). The structured VNF-FG for request $R$ is represented as a tree [4] in which leaf nodes represent VNFs belonging to $vnf_R$, while a middle node with index $i$, namely $S_i$, represents one of the sub-structures i.e., $S_i \in \{seq, par, sel, loop\}$. Each middle node $S_i$ has at least two children where a child can be either a VNF or a sub-structure (See Fig. 1). According to the constructed tree, we can define the relation between VNFs in the chain; the immediate predecessor of a VNF $f^t$ can be determined by parsing the tree [4]. Let $IP(f^t)$ denote the immediate predecessor of $f^t$ and $(IP(f^t), f^t)$ a VNF edge if and only if the packets from VNF $IP(f^t)$ must be forwarded to the VNF $f^t$. We assume that the amount of traffic sent from $IP(f^t)$ to $f^t$ for request $R$ is $A_{f^t}^R$. More details of the structured VNF-FG can be found in [4].

Table I. Some of the notations and decision variables used in this paper

| Input Parameters | |
|---|---|
| $f^t$ | VNF of type $t \in T$ |
| $\vartheta_{f^t}$ | Resource requirements for processing $f^t$ (in processing units) |
| $c_{f^t}$ | Processing capacity of $f^t$ (in traffic units) |
| $I_{f^t}$ | Set of VNF instances associated to $f^t$ |
| $\partial_{f^t}$ | License cost for $f^t$ |
| $Req$ | Set of structured VNF-FG requests assigned to the system |
| $vnf_R$ | Set of required VNF types for request $R$, $vnf_R \subset T$ |
| $A_{f^t}^R$ | Amount of traffic from $IP(f^t)$ to $f^t$ for request $R$ |
| $N^Z$ | Set of cloud/fog nodes |
| $U_R$ | Set of IoT/end-users for request $R$ |
| $E$ | Set of all possible communication in the network |
| $c_n$ | Cloud/fog node capacity (in processing resource units) |
| $\gamma_n$ | Cloud/fog node cost per processing unit usage |
| $D_n^{f^t}$ | The processing delay of $f^t$ on node $n$ |

| Input Parameters (Cont.) | |
|---|---|
| $D_{e_{nm}}(A, X, Y)$ | The transmission delay of sending traffic $A$ through $e_{nm}$ |
| $\rho_{e_{nm}}(A, X, Y)$ | The transmission cost of sending traffic $A$ through $e_{nm}$ |
| $BW_{e_{nm}}(X, Y)$ | The bandwidth capacity of $e_{nm}$ |
| $Lat_{e_{nm}}(X, Y)$ | The network latency of $e_{nm}$ |
| $v^n$ | Movement velocity of node $n$ |
| $p_{st}^n$ | Probability that node $n$ is static |
| $p_p^n$ | Probability that a node $n$ is in pause |
| $f_x^n(X)$ | Stationary pdf of location $X = (X, Y)$ of node $n$ |
| $f_{init}^n(X)$ | Initial spatial distribution of location $X = (X, Y)$ of node $n$ |
| **Decision Variables** | |
| $x_{i,f^t,n}$ | Binary variable, indicating if instance $i$ of VNF type $t$ is instantiated on cloud/fog node |
| $x_{i,f^t,n}^R$ | Binary variable, indicating if instance $i$ of VNF type $t$ instantiated on cloud/fog node is assigned to request $R$ |

**Network** – We consider $N^Z$ as a set of cloud and fog nodes, where $Z = C|F$ is used to indicate cloud or fog. We use $c_n$ and $\gamma_n$ to represent the capacity and the cost respectively, per unit of resource (e.g., CPU, memory, storage) usage of node $n \in N^Z$. We represent the threshold for resource usage of a cloud/fog node with $\mu_n$. The delay per traffic unit processing of VNF type $t$ hosted on a cloud/fog node $n^n \in N^Z$ is represented by $D_n^{f^t}$. The application components may communicate with IoT/end-users. We denote the set of IoT/end-users for request $R \in Req$ by $U_R$.

Given a network of cloud nodes, fog nodes, and IoT/end-user devices, we consider $E$ to be the set of all possible communications in the network, then an edge is represented by $e_{nm} \in E$, which represents communication between any two cloud/fog nodes, or one cloud node and a fog node, or any IoT/end-users and cloud/fog nodes. Note that node indices are $n$ and $m$ in this notation. When the location of $n$ is $X$ and the location of $m$ is $Y$, then, for $e_{nm}$, we define $D_{e_{nm}}(A, X, Y)$ and $\rho_{e_{nm}}(A, X, Y)$, that represent the delay and the cost, respectively, of transmitting traffic amount of $A$ through $e_{nm} \in E$. For the communications between the VNFs and IoT/end-users, we define two matrices, $\widetilde{\omega}_{l \times k}^R$ and $\tilde{A}_{l \times k}^R$ that represent respectively the communication and the amount of traffic exchanged between IoT/end-users and the VNFs of request $R$. Here, $l$ represents the number of IoT/end-users communicating with VNFs in request $R$ while $k$ represents the number of VNFs in $R$. $\omega_{uf^t}^R \in \{0,1\}$ is 1 if there is communication between IoT/end-user $u$ and the VNF $f^t$ of request $R$, while $A_{uf^t}^R$ provides the amount of traffic exchanged between IoT/end-user $u$ and the VNF $f^t$ of request $R$. We also define $BW_{e_{nm}}(X, Y)$ represents the bandwidth capacity of $e_{nm}$, and $Lat_{e_{nm}}(X, Y)$ represents the network latency when node $n$ is in location $X$ and node $m$ is in location $Y$. In this regard, the same nodes will communicate with different delays, costs, and bandwidths when they are located in various locations. We represent the threshold for usage of the bandwidth capacity by $\mu_{e_{nm}}$.

We assume that cloud nodes, fog nodes, and IoT/end-users are located in a two-dimensional rectangular region $Q \in [0,1]^2$. Note that two-dimensional localization has also been used in ad-hoc networks [32]. Thus, $X = (x, y) \in [0,1]^2$ denotes the location of a cloud/fog node or an IoT/end-user device. The locations of IoT/end-users are assumed to be fixed and defined in the region $Q$.

## 4. CLOUD/FOG NODE LOCATION ANALYSIS AND OPTIMIZATION FORMULATION

Here we first calculate the Probability Density Function (PDF) of cloud/fog nodes locations, and then we explain the objective function and the constraints of our optimization model.

### 4.1. Cloud/Fog Node Location Analysis

In this section, we first explain the Random Waypoint Model (RWP) that we have used to model the mobility of a fog node, then, we explain the analysis provided by [17] to calculate the distribution of a node location when it moves according to RWP. Note that this model can also be used for static fog nodes and cloud nodes as will be discussed at the end of this section.

According to RWP, nodes move in a square region of $Q \in [0,1]^2$ independently of each other. The whole movement trace is defined as the repetition of some movement trajectories; each trajectory represents a movement vector from a source to a destination that is performed in a period. Indeed, at each period, node $n$ selects a destination location in region $Q$; moves toward it with velocity $v^n$; and pauses there for a random duration. The new period is then started from the destination location of the previous period. The authors of

[17] have provided a detailed analysis of RWP when the movement process is repeated for infinite time; in other words when the movement process becomes stationary.

Now, we give the analysis in [17] that calculates the location distribution of a node when it moves according to the RWP model. The location of a node is modeled as a random variable namely, $X$ with a possible value denoting a point in the region $Q$ i.e., $X = (x, y)$. The aim is to calculate $f_X^n(X) = f_X^n(x,y)$ which is the probability that the fog node $n$ is in location $X = (x, y)$. Before proceeding to the calculation, we introduce some notations used in [17]. Let $f_{init}^n(X)$ be the distribution according to which node $n$ initially is located on region $Q$. Let $p_{st}^n$ be the probability that a node is static. This parameter has been considered in order to include the fraction of nodes that may not move in the model (such as static fog nodes or cloud nodes in this paper). The random variable for pause time is represented by $PS$. In this regard, $f_{PS}^n(ps)$ is the probability that pause time equals to $ps$ within a period. We assume that the expected value of the distribution i.e., $E[PS]$ is known. Let $L$ be the random variable defining the trajectory length. In RWP, it can be shown that the expected trajectory length is equivalent to the expected distance between two independent points chosen uniformly at random in region $Q \in [0,1]^2$, which is calculated as $E(L) = 0.52$ [17][33]. For mathematical details, readers are referred to [33][34].

Now, we explain the calculation of node location distribution i.e., $f_X^n(X)$ as shown in (1) [17]. The calculation includes the summation of three terms that respectively are related to static, pause, and mobility contributions in location probability calculation. Regarding the first term, when the node is static, the initial PDF i.e., $f_{init}^n(X)$ defines the probability of node location at point $X = (x, y)$. The second term i.e., $(1 - p_{st}^n) \cdot p_p^n$ indicates the probability that the node is not static, but it is in pause at location $X$. Here, $p_p^n$ is the pause probability and is calculated as the fraction of time that the node is in pause as calculated by (2). Note that in (2), $\frac{1}{v^n} E(L)$ calculates the expected duration that the node is mobile. Finally, the third term in (1) is for the case that the node is not static and does move. In this case, the probability that the node is in location $X$ is calculated by the distribution $f_m^n(X)$.

$$f_X^n(X) = p_{st}^n f_{init}^n(X) + (1 - p_{st}^n)p_p^n + (1 - p_{st}^n)(1 - p_p^n)f_m^n(X) \qquad (1)$$

$$p_p^n = \frac{E(PS)}{E(PS) + \frac{1}{v^n}E(L)} \qquad (2)$$

Now, we explain the calculation of $f_m^n(X)$. Regarding the symmetries in the square shape of $Q$, the whole region can be divided into eight triangular sub-regions. To calculate the $f_m^n(X)$, it is sufficient to know the distribution in the sub-region $Q^* = \{(x, y) \in [0,1]^2 | (0 < x \le 0.5), (0 < y \le x)\}$. The distribution in other sub-regions of $Q$ is obtained by variable substitution [17]. The probability of being located at location $X = (x, y)$ inside $Q^*$ is denoted by $f_m^*$; it is approximated by calculating the probability that the node is located in a square with sufficiently small length centered at $(x, y)$. This probability depends on the source of the movement trajectory and $X$ (the location). Through geometrical analysis, the probability is calculated as the aggregation of convex polygons generated by lines crossing the trajectory source and corners of the square (for all possible sources). Eq. (3) shows the close form calculation of $f_m^*$. Readers are referred to [17] for the mathematical details.

$$f_m^*(x,y) = 6y + \frac{3}{4}(1 - 2x + 2x^2)\left(\frac{y}{y-1} + \frac{y^2}{(x-1)x}\right) + \frac{3y}{2}\left[(2x-1)(y+1)\ln\left(\frac{1-x}{x}\right) + (1 - 2x + 2x^2 + y)\ln\left(\frac{1-y}{y}\right)\right] \qquad (3)$$

Eq. (4) shows the distribution in all sub-regions of $Q$ based on $f_m^*$ [17]. Note that after calculating the mobility probability contribution i.e., $f_m^n$, the probability of node location can be calculated by (1) as it is the target of the analysis.

$$f_m^n(X) = f_m^n(x,y) = \begin{cases} f_m^*(x,y) & 0 < x \le 0.5, 0 < y \le x \\ f_m^*(y,x) & 0 < x \le 0.5, x \le y \le 0.5 \\ f_m^*(1-y,x) & 0 < x \le 0.5, 0.5 \le y \le 1-x \\ f_m^*(x,1-y) & 0 < x \le 0.5, 1-x < y < 1 \\ f_m^*(1-x,y) & 0.5 \le x < 1, 0 < y \le 1-x \\ f_m^*(y,1-x) & 0.5 \le x < 1, 1-x \le y \le 0.5 \\ f_m^*(1-y,1-x) & 0.5 \le x < 1, 0.5 \le y \le x \\ f_m^*(1-x,1-y) & 0.5 \le x < 1, x \le y < 1 \\ 0 & \text{otherwise} \end{cases} \qquad (4)$$

Though cloud nodes are static and do not require analysis, we use Eq. (1) for cloud nodes as well, to be able to provide the formulas in the next subsection in a general manner. Note that the PDF calculated in Eq.

(1) and the performed mobility analysis can be applicable to both static fog nodes and cloud nodes. For instance, considering cloud nodes, $p_{st}^n$ becomes 1; using Eq. (1), the distribution of a cloud node's location becomes the same as its initial location i.e., $f_x^n(X) = f_{init}^n(X)$. More precisely, the cloud node location distribution is a probability mass function with the value of 1 for the location of the cloud node.

*4.2. Optimization Formulation*

We formulate our problem as an optimization problem with the objective of minimizing the weighted aggregated function of makespan (graph completion time) and cost of all requests. We define the following decision variables:

$$x_{i,f^t,n} = \begin{cases} 1, & \text{if instance } i \text{ of } f^t \text{ is placed on } n \\ 0, & \text{otherwise} \end{cases} \quad (5)$$

$$x_{i,f^t,n}^R = \begin{cases} 1, & \text{if instance } i \text{ of } f^t \text{ on } n \text{ is assigned to request } R \\ 0, & \text{otherwise} \end{cases} \quad (6)$$

In the rest of this section, we explain the expected makespan and cost calculations, followed by the objective function and the constraints. Table I lists the key notations and decision variables.

*4.2.1. Makespan and Communication Cost Computation*

The makespan is an application's execution time, defined as the time it takes for the first component to start execution until the execution of the last component is completed [35] [4]. Note that the communication times with the IoT/end-users are also included in makespan calculations. In turn, the application execution cost is defined as the monetary cost for communication between application components and also between IoT/end-users and components.

The calculation of the expected application makespan and communication cost is performed based on parsing the associated tree structure of the VNF-FG, as explained in Section 3. The time and cost of the leaf nodes representing the VNFs are calculated first. These values are then aggregated to calculate the time and the cost for the middle nodes. The middle nodes represent sub-structures. The total makespan and the cost of the root of the tree are then calculated by aggregating the calculated time/cost values of the nodes from the bottom to the top according to the tree structure.

*a. VNFs-Level Calculation*

The processing time of the traffic received by each VNF from its immediate predecessors belonging to request $R$ is calculated as below:

$$M_{proc}(R, f^t) = \sum_{n \in N^Z} \sum_{i \in I_{f^t}} x^R_{i,f^t,n} \cdot A^R_{f^t} \cdot D^{f^t}_n \qquad (7)$$

The communication time required to transmit traffic $A^R_{f^t}$ to a VNF $f^t$ belonging to a VNF-FG request $R$ from $IP(f^t)$ and to transmit traffic $A^R_{uf^t}$ between $f^t$ and IoT/end-users is calculated as (8). Here, $E(D_{e_{nm}}(A^R_{f^t}))$ is the expected delay of transmitting traffic $A^R_{f^t}$ on link $e_{nm}$. Similarly, $E(D_{e_{nu}}(A^R_{uf^t}))$ is the expected delay for the transmission of $A^R_{uf^t}$ amount of traffic between an IoT/end-user and a cloud/fog node.

$$M_{com}(R, f^t) = \max(\sum_{n,m \in N^Z} \sum_{i,j \in I_{f^t}} x^R_{i,f^t,n} \cdot x^R_{j,IP(f^t),m} \cdot E(D_{e_{nm}}(A^R_{f^t})),$$
$$\sum_{n \in N^Z} \sum_{i \in I_{f^t}} \sum_{u \in U_R} x^R_{i,f^t,n} \cdot \omega^R_{uf^t} \cdot E(D_{e_{nu}}(A^R_{uf^t}))) \qquad (8)$$

The expected delay of transmitting traffic size $A$ through the link $e_{nm}$, is calculated by (9). Here, the expected delay is calculated as an aggregation of transmission delay over possible geographical distributions of the nodes $n$ and $m$. The geographical distribution probabilities are respectively $f^n_x(X)$ and $f^m_y(Y)$ that can be calculated by (1). Note that in (9), $D_{e_{nm}}(A, X, Y)$ is the data transfer time for sending traffic $A$ on edge $e_{nm}$ when nodes $n$ and $m$ are respectively in locations $X$ and $Y$. It is calculated as the summation of network latency and the relation of the size of the traffic to the link bandwidth. This calculation has been given in (10).

$$E(D_{e_{nm}}(A)) = \int_{[0,1]^2} \int_{[0,1]^2} f^n_x(X) \cdot f^m_y(Y) \cdot D_{e_{nm}}(A, X, Y) \, dX \, dY \qquad (9)$$

$$D_{e_{nm}}(A, X, Y) = \frac{A}{BW_{e_{nm}}(X, Y)} + Lat_{e_{nm}}(X, Y) \qquad (10)$$

The expected delay for the transmission of $A$ amount of traffic between an IoT/end-user (in location $Z$) and the cloud/fog node $n$ (in location $X$) is calculated by (11). The delay is calculated as an aggregation of the delays over possible geographical distribution of node $n$ defined by $f^n_x(X)$.

Table II. The cost and the makespan estimation for $S_i \in \{seq, par, sel, loop\}$

| Sub-structures | Communication Cost $C_{com}(S_i)$ | Processing Time $M_{proc}(S_i)$ | Communication Time $M_{com}(S_i)$ |
|---|---|---|---|
| $S_i$ is seq | $\sum_{f^t \in S_i} C_{com}(f^t)$ | $\sum_{f^t \in S_i} M_{proc}(f^t)$ | $\sum_{f^t \in S_i} M_{com}(f^t)$ |
| $S_i$ is par | $\sum_{f^t \in S_i} C_{com}(f^t)$ | $\max_{f^t \in S_i} M_{proc}(f^t)$ | $\max_{f^t \in S_i} M_{com}(f^t)$ |
| $S_i$ is sel | $\sum_{f^t \in S_i} h_{f^t} \cdot C_{com}(f^t)$ | $\sum_{f^t \in S_i} h_{f^t} \cdot M_{proc}(f^t)$ | $\sum_{f^t \in S_i} h_{f^t} \cdot M_{com}(f^t)$ |
| $S_i$ is loop | $it. \sum_{f^t \in S_i} C_{com}(f^t)$ | $it. \sum_{f^t \in S_i} M_{proc}(f^t)$ | $it. \sum_{f^t \in S_i} M_{com}(f^t)$ |

$$E\left(D_{e_{nu}}(A)\right) = \int_{[0,1]^2} f_x^n(\boldsymbol{X}) \cdot D_{e_{nu}}(A, \boldsymbol{X}, \boldsymbol{Z}) \, dX \qquad (11)$$

The approach described above for time can also be used to calculate the communication costs which is the bandwidth cost incurred by utilizing the links. When a VNF like $f^t$ runs on a node like $n$, it receives traffic from its immediate predecessor that may run on a node like $m$ (through the link $e_{nm}$). The expected cost is $E(\rho_{e_{nm}}(A_{f^t}^R))$. Similarly, traffic from IoT/end-users $u$ to $f^t$ may be transmitted through the link $e_{nu}$. The expected transmission cost is $E(\rho_{e_{nu}}(A_{uf^t}^R))$. In this regard, the communication cost is calculated as the sum of those mentioned costs as calculated by (12). Note that to calculate $E(\rho_{e_{nm}}(A_{f^t}^R))$, the $D_{e_{nm}}(A, \boldsymbol{X}, \boldsymbol{Y})$ in Eq. (9) should be replaced by $\rho_{e_{nm}}(A, \boldsymbol{X}, \boldsymbol{Y})$. Similarly, to calculate $E(\rho_{e_{nu}}(A_{uf^t}^R))$, the $D_{e_{nu}}(A, \boldsymbol{X}, \boldsymbol{Z})$ in Eq. (11) must be replaced by $\rho_{e_{nu}}(A, \boldsymbol{X}, \boldsymbol{Z})$.

$$C_{com}(R, f^t) = \sum_{n \in N^Z} \sum_{i,j \in I_{f^t}} x_{i,f^t,n}^R \cdot x_{j,IP(f^t),m}^R \cdot E(\rho_{e_{nm}}(A_{f^t}^R)) + \\ \sum_{n \in N^Z} \sum_{i \in I_{f^t}} \sum_{u \in U_R} x_{i,f^t,n}^R \cdot \omega_{uf^t}^R \cdot E(\rho_{e_{nu}}(A_{uf^t}^R)) \qquad (12)$$

*b. VNF-FG Level Calculation*

The calculations of the processing/communication time and the communication cost for the sub-structures sequence, parallel, selection, and loop are shown in Table II. In a sequence sub-structure, the time and the cost of all of its children are accumulated. A loop can be considered as a sequence structure that is repeated for a certain number of iterations. We define $it$ as the expected number of iterations of a loop structure, it is calculated as: $it = \frac{q}{1-q}$ where $q$ is the probability of the loop's occurrence. For a parallel sub-structure, all of its children are executed in parallel, hence the time is determined based on the maximum time value of its children. However, the cost is the sum of the costs for all children. The calculation for a selection sub-structure, the probabilities of the selection's children are involved in the calculation. Let $h_{f^t}$ represent the probability of selecting a child $f^t$ of a selection sub-structure. $h_{f^t} = 1$ for the children of sequence, parallel, and loop sub-structures.

Finally, to calculate the total makespan and the cost of a VNF-FG, the makespan and the cost of the root of the tree are computed by aggregating the time and the cost of the VNFs and of the basic sub-structures in a bottom-to-top manner according to the tree structure. The total makespan and the total cost of a VNF-FG request $R$ are calculated as given in Eq. (13) and Eq. (14), respectively:

$$M(R) = M_{proc}(R, root) + M_{com}(R, root) \quad (13)$$

$$C(R) = C_{com}(R, root) \quad (14)$$

### 4.2.2. Optimization Formulation

In this section, we explain the objective function and the constraints of the optimization problem. Our objective is to enable the embedding of VNF-FGs in cloud and fog NFVIs such that the makespan and the cost are minimized, as shown in Eq. (15).

$$obj = Min \left( \alpha \sum_{\forall R \in Req} M(R) + (1 - \alpha) \left[ \sum_{\forall R \in Req} C(R) + C_{dep} \right] \right) \quad (15)$$

The deployment cost; $C_{dep}$, represents both the license cost of VNFs and the hosting cost,

$$C_{dep} = C_{Lic} + C_{hst} \quad (16)$$

The license cost is the cost of the total software license costs for the VNFs instantiation,

$$C_{Lic} = \sum_{n \in N^Z} \sum_{t \in T} \sum_{i \in I_{f^t}} x_{i,f^t,n} \cdot \partial_{f^t} \quad (17)$$

and the hosting cost is the cost of the assigned resources (e.g., CPU, memory, storage) to VNFs belonging to VNF-FG requests. It is calculated as:

$$C_{hst} = \sum_{n \in N^Z} \sum_{i \in I_{f^t}} x_{i,f^t,n} \cdot \gamma_n \cdot \vartheta_{f^t} \quad (18)$$

In Eq. (15), $\alpha$ is the weight parameter that defines priorities between makespan and cost, $1 \geq \alpha \geq 0$. $\alpha = 1$ motivates placement in the fog, while $\alpha = 0$ motivates placement in the cloud. Generally, the fog provides lower latency due to its proximity to IoT/end-users, however, the resources in the fog are more expensive.

Now, we explain the constraints involved in the problem. Eq. (19) ensures that the total resources required by instances of all VNF types do not exceed the capacity of a cloud/fog node:

$$\sum_{t \in T} \sum_{i \in I_{f^t}} \vartheta_{f^t} \cdot x_{i,f^t,n} \leq \mu_n \cdot c_n \quad \forall n \in N^Z \tag{19}$$

Eq. (20) ensures that the communication links where the source and the destination are both in the cloud or both in the fog, or where the source is in one and the destination is in the other are not overloaded from the aspect of link utilization. A similar discussion exists for the communication links between the IoT/end-users and cloud/fog nodes according to the constraint in Eq. (21).

$$\sum_{\forall R \in Req} A_{f^t}^R \cdot x_{i,f^t,n}^R \cdot x_{j,IP(f^t),m}^R \leq \mu_{e_{nm}} \cdot BW_{e_{nm}} \tag{20}$$
$$\forall n, m \in N^Z$$

$$\sum_{\forall R \in Req} A_{uf^t}^R \cdot x_{i,f^t,n}^R \cdot \omega_{uf^t}^R \leq \mu_{e_{nu}} \cdot BW_{e_{nu}} \tag{21}$$
$$\forall n \in N^Z, u \in U_R$$

Eq. (22) ensures that the capacity of an instance of a VNF $f^t$ is not exceeded by the total traffic requested by its immediate predecessor(s) and the IoT/end-users communicating with it.

$$\sum_{\forall R \in Req} (A_{f^t}^R \cdot x_{i,f^t,n}^R + A_{uf^t}^R \cdot x_{i,f^t,n}^R \cdot \omega_{uf^t}^R) \leq \mu_{f^t} \cdot c_{f^t} \tag{22}$$
$$\forall t \in T, \forall i \in I_{f^t}, \forall u \in U_R, \forall n \in N^Z$$

Eq. (23) ensures that the assigned VNF instances are already deployed in the network and Eq. (24) ensures that at least one instance of each required VNF type is deployed.

$$x_{i,f^t,n}^R \leq x_{i,f^t,n} \tag{23}$$
$$\forall R \in Req, f^t \in vnf_R, i \in I_{f^t}, n \in N^Z$$

$$\sum_{\forall n \in N^Z} \sum_{\forall i \in I_{f^t}} x_{i,f^t,n} \geq 1 \quad \forall t \in T \tag{24}$$

It should be noted that Eq. (8) and (20) and the processing and communication time equations in Table II for parallel sub-structure are non-linear. However, they can be linearized. Readers are referred to [4] for linearization techniques that can be applied to the non-linear equations in this paper.

Note that for $|T|$ VNF type, $I$ VNF instances for each type, $|Req|$ requests, and $|N^z|$ cloud/fog nodes the search space size is of $O(2^{|T|.I.|Req|.|N^z|})$. Finding the exact solution for such exponential space takes extensive time. Each possible placement has its own makespan/cost. Thus, an efficient heuristic is required to solve the problem and perform the appropriate trade-off between the makespan and the cost. Note that the place of the components on cloud/fog nodes can influence both makespan and cost.

## 5. TABU SEARCH-BASED COMPONENT PLACEMENT

In this section, we propose a Tabu Search-based Component Placement (TSCP) algorithm for the optimization problem explained in Section 4. The search space size is exponential in terms of the number of VNF types, VNFs instances, number of cloud/fog nodes, and number of requests. Thus, as will be seen in Section 6, the run time for finding the optimal solution with CPLEX is quite long, even for small-scale scenarios. Therefore, a heuristic approach is required to make the placement in real system scales with acceptable run times. Tabu Search meta-heuristic has been shown to be promising in terms of finding a near-optimal solution in combinatorial optimization problems (e.g., [33][34]) and VNF placement problems [38][39], and so we exploit it in our component placement algorithm.

Tabu is an iterative search process that starts exploring the search space from an initial solution and iteratively performs moves to transit from the current solution to a better one in its neighborhood until the stopping criterion is satisfied. Tabu Search uses a memory structure called Tabu-list to avoid looping during the search process, thereby preventing cycling to previously visited solutions [40]. In the rest of this section, we explain the major elements of the Tabu Search algorithm as outlined in Algorithm 1.

1. Tabu starts searching with an initial placement. VNF types that communicate with IoT devices are randomly assigned to fog nodes with enough capacity to process the VNF. The rest of the VNFs are assigned randomly to cloud nodes with sufficient capacity Eq. (19). Note that the constraints satisfaction in the search process will be considered in the evaluation phase as will be discussed later in this section.

2. Tabu explores the neighborhood of the current placement to improve the quality of the just-identified best placement. A neighborhood is generated by applying a single move from the current placement. We define four moves as below:

   **VNF Reassignment** – A VNF is selected randomly and moved to a node with enough capacity and minimum amount of aggregated processing time, hosting cost, and communication time/cost with its immediate predecessors and IoT/end-users (for all the requests using this VNF). Note that the aggregation is performed as in the weighting used in Eq. (15).

   **Bulk VNF reassignment -** A node is selected randomly and the VNFs on it are assigned to another node with enough capacity to host the VNFs and minimum amount of aggregated processing time, hosting cost, and communication time/cost with its immediate predecessors and IoT/end-users for all VNFs.

   **Request reassignment-** A request is selected randomly and one of its required VNFs is assigned to another instance with enough capacity to tolerate the traffic and minimum amount of aggregated processing time, hosting cost, and communication time/cost with its immediate predecessors and IoT/end-users.

   **Bulk request reassignment -** A VNF is selected, and all its requests are assigned to another VNF instance with enough capacity and minimum amount of aggregated processing time, hosting cost, and communication time/cost with its immediate predecessors and IoT/end-users.

3. To avoid visiting the same solution several times, Tabu uses a list called Tabu list to store moves marked as Tabu. The move that generates the best neighborhood i.e., $best\_move$ is saved in $Tabu\_list$ for a specific length of time, or number of iterations i.e., $i_{tabu}$. Further, a Tabu move can be released from $Tabu\_list$ if it meets the aspiration criterion, defined as the case when a better solution than the current best solution has been found.

| | Algorithm 1: Tabu Search Algorithm |
|---|---|
| 1 | **initialization:** Create initial placement randomly $S_0$, |
| 2 | $S_{curr} \leftarrow S_0, S_{best} \leftarrow S_0, j \leftarrow 0, Tabu\_list \leftarrow \emptyset$ |
| 3 | **while** $j \leq i_{stop}$ |
| 4 |    $neighborhood\_list \leftarrow$ create candidate neighborhood list |
| 5 |    **for** each $neighbor \in [neighborhood\_list]$ |
| 6 |       evaluate the neighbor $E(neighbor)$ |
| 7 |    **end** |
| 8 |    $best\_neighbor \leftarrow \underset{Neighbor}{\mathrm{argmin}}\, E(neighbor)$ |
| 9 |    $best\_move \leftarrow$ select the move that led to $best\_neighbor$ |
| 10 |    $j \leftarrow j + 1$ |
| 11 |    **if** $best\_move$ is not in $Tabu\_list$ |
| 12 |       $Tabu\_list \leftarrow best\_move$ for $i_{tabu}$ iterations |
| 13 |    **else if** $fitness(best\_neighbor) < E(S_{best})$ |
| 14 |       remove $best\_move$ from $Tabu\_list$ |
| 15 |    **end** |
| 16 |    **if** $fitness(best\_neighbor) < E(S_{best})$ |
| 17 |       $S_{best} \leftarrow best\_neighbor$ |
| 18 |       $j \leftarrow 0$ |
| 19 |    **end** |
| 20 | **end** |
| 21 | **return** $S_{best}$ |

4. In each iteration of the Tabu search process, the neighbors are evaluated in order to recognize the best solution and move towards that. We use the aggregation of the objective function as defined by Eq. (15) and the penalty function imposed due to constraints' violation to evaluate each placement. Eq. (25) indicates the evaluation function:

$$E(S_{curr}) = \begin{cases} obj(S_{curr}), & \text{If } S_{curr} \text{ is feasible} \\ obj(S_{curr}) + p(S_{curr}), & \text{otherwise} \end{cases} \quad (25)$$

where $p(S_{curr})$ is the penalty function for the current placement. We have used the suggested penalty calculation in [41]. The left and right sides of the constraints (19), (20), (21), and (22) are represented with $g_m$ for $m = 1 \ldots 4$, and $b_m$ respectively. In this regard, a constraint can be represented by $g_m(S_{curr}) < b_m$. The penalty is calculated as below:

$$p(S_{curr}) = \sum_{m=1}^{M} \varsigma_m \max(0, g_m(S_{curr}) - b_m) \quad (26)$$

$\varsigma_m$ is the normalization coefficient to make $p(S_{curr})$ and $obj(S_{curr})$ in the same scale.

5. The algorithm will stop when the best solution (i.e., $S_{best}$) does not improve for a certain number of consecutive iterations ($i_{stop}$).

## 6. PERFORMANCE EVALUATION

Here we evaluate the performance of our proposed placement algorithm, the TSCP, comparing it with the optimal solution gained by CPLEX (Optimal), to the TSCP (Random Explore) where the optimization variables are changed by random moves instead of makespan/cost driven moves as discussed in Section 5, and finally, to a first-fit greedy placement (Greedy). Greedy iterates over the set of VNF-FGs associated with applications. For each VNF in a VNF-FG, Greedy first checks if that VNF type is already deployed in the network and if it has adequate capacity. If such a deployed VNF is found, Greedy assigns it to the request. Otherwise, Greedy instantiates a new VNF of that type on the first fitted node (from the aspect of VNF processing and communication with the immediate predecessors). In the rest of this section, we explain the experimental setup and then we present the evaluation results.

### 6.1. Simulation Setup

**VNFs -** We have assumed a license cost of $100 for the VNF instantiation. Each VNF uses an OpenStack VM from tiny to large size, with 1 to 4 vCPUs.

**VNF-FGs** - We have synthesized 50 loop-free VNF-FGs with the structure of fork-join, using the method proposed in [42]. For each graph, the number of VNFs has been chosen randomly between 3 and 10. The height of the graph that is equal to the length of the longest path from the root to the last VNF, is chosen randomly from {2, 4, 6, 8}. Each graph has been built in two steps: 1) regarding the number of VNFs and height, an equal out-degree is assigned for nodes at which the fork (split) happens. 2) For the purpose of randomization, a post-process is performed on the graph structure. Indeed, the parameter edge ratio that is randomly chosen from {1.1, 1.3, 1.5, 1.7, 1.9}, is used as a multiplier of nodes out-degree to increases the out-degree of the nodes randomly; accordingly, the number of split nodes is tuned randomly to generate a graph with the specified number of VNFs. For each fork (split), the selection to the parallel ratio that indicates the occurrence probability of selection rather than parallel is randomly chosen from {0, 0.2, 0.4, 0.6, 0.8, 1}. For the sake of simplicity, equal probabilities are assigned to the children of selection sub-structure. We consider the size of the data transmitted in the chains is selected randomly in the range of 100 bytes to 80 KB [43].

Table III. Summary of simulation parameters

| Parameter | Value | | Parameter (Cont.) | Value |
|---|---|---|---|---|
| *VNFs* | | | Number of IoT/end-users | [5-30] |
| Number of VNF types | [3-27] | | Bandwidth cost ($/GB): cloud, fog, cloud-fog, IoT-cloud, IoT-fog | 0.155, [0.25-2], [10-20], 20, [0.05-0.25] |
| VNF resource requirements (vCPU) | [1- 4] | | Bandwidth (Gbps): cloud, fog, cloud-fog, IoT-cloud and IoT-fog | 10, [0.1-1], [1, 10], 10, [250Kbps-54Mbps] |
| VNF processing capacity per GB | [1- 2] | | | |
| VNF license cost ($) | 100 | | Latency (msec): cloud, fog, cloud-fog, IoT-cloud and IoT-fog | [50-100], [10-50], [100-255], 250, [7-20] |
| *VNF-FGs* | | | | |
| Number of VNF-FG requests | [1-50] | | *Cloud/Fog Nodes* | |
| Number of VNFs in a VNF-FG request | [3-10] | | Nodes capacity (vCPU): cloud, fog | 8, [2-4] |
| Traffic Amount (KB) | [0.1-180] | | Nodes cost ($/vCPU): Cloud, fog | [2.33- 4.65], [4.65- 5.82] |
| *Network* | | | Nodes delay (msec/MB): cloud, fog | 0.25, 25 |
| Number of nodes: cloud, fog | [2, 4, 8], [3, 6, 12] | | | |

**Network topology** – To the best of our knowledge, currently there is no real cloud/fog infrastructure or testbed to inspire the topology. However, in order to validate our proposed algorithm, we have used similar parameters and topologies as in the literature (e.g., [26], [6]). The experiments have been done on two network topologies. The first topology consists of 10 nodes including 4 cloud nodes and 6 fog nodes, and the second topology consists of 20 nodes including 8 cloud nodes and 12 fog nodes. The number of IoT/end-users ranges from 5 to 30 per application. We assume logical links exist between each pair of cloud/fog nodes. where the link bandwidth capacity between cloud nodes is 10Gbps, between fog nodes taken randomly in the range of 100Mbps to 1Gbps, and between cloud and fog nodes are taken randomly in the range of 1Gbps to 10Gbps [43]. The bandwidth cost for links between cloud nodes is $0.155 per GB transmission, for links between fog nodes is taken randomly between $0.25 and $2 per GB transmission, and for the links between cloud and fog nodes is taken randomly in the range of $10 to $20 per GB transmission [39]. The communication latency (propagation delay) between nodes in the cloud, in the fog, and between the cloud and the fog nodes ranges within (50 to 100) msec, (10 to 50) msec, and (100 to 255) msec, respectively [6][26].

Similarly, we assume logical links exist between each pair of cloud/fog node and IoT devices. The communication bandwidth between IoT devices and the cloud is 10Gbps, whilst it is in the range of 250Kbps to 54Mbps for communication with fog nodes [43]. The bandwidth cost for the links between IoT devices and cloud and fog nodes are set to $20/GB and uniform in the range of $(0.05 to 0.25)/GB, respectively [44]. The communication latencies between IoT devices and the cloud and the fog nodes are set to 250msec, and in the range of 7 to 20 msec, respectively. Note that the bandwidth, the cost, and the latency in communications vary randomly in the mentioned ranges depending on the fog node location involved in the communication.

**Cloud/fog nodes -** We consider the cloud nodes have 8 vCPUs and the fog nodes have a random number between 2 and 4 vCPUs. The cost of cloud/fog node usage is selected randomly in the range of $(2.33 to 4.65)/vCPU and $(4.65 to 5.82)/vCPU respectively [45]. The processing delay on cloud and fog nodes is set to 0.25 msec and 25 msec, respectively, per Megabyte traffic processing [43].

Finally, we found the values of 60 and 20 appropriate for the experiments for $i_{tabu}$ and $i_{stop}$, respectively. For all the experiments, we assume that the whole capacity of the VNFs and communication links can be used. All simulations were conducted on a single machine with dual 2X8-Core 2.50GHz Intel Xeon CPU E5-2450v2 and 40GB memory. Table III lists the parameters in the simulation.

*6.2. Evaluation Results*

In the rest of this section, the average of the normalized cost, makespan, and aggregated of them for all the requests are given for 10 runs. To assess the effect of mobility consideration in application placement, Fig. 2 illustrates the aggregated makespan/cost improvement of TSCP in comparison with our previous work, namely called Placement with the assumption of Static Fog nodes (PSF) [4]. Note that the improvement ratio has been calculated as the division of the aggregated value in PSF by the aggregated value in TSCP. Here, $\alpha = 0.5$ and the probability that a fog node is mobile i.e., $1 - p_{st}^n$ changes in the range of 0, 0.25, …1. As it can be seen, TSCP performs the same as PSF when all fog nodes are static. However, it outperforms the PSF for all other probabilities. The more the probability of mobility, the more is the outperformance. The reason is that PSF utilizes only the initial places of the fog nodes to decide for the placement, whilst TSCP considers the mobility pattern of the nodes in the placement. As expected, this consideration becomes more significant when more fog nodes become mobile.

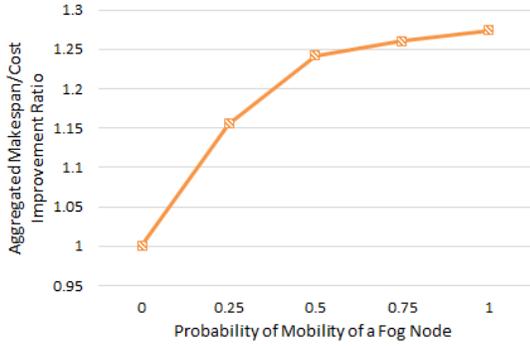
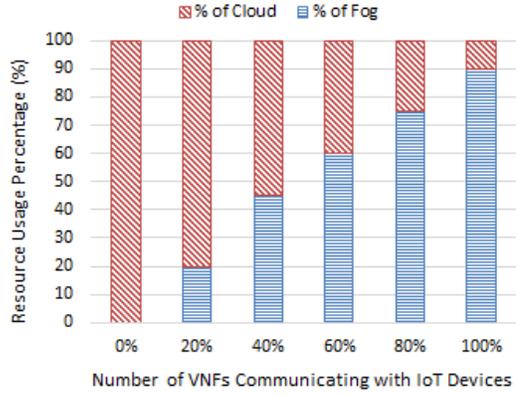

Fig. 2. Outperformance of TSCP in comparison with PSF for 10 nodes and up to 15 VNF-FG requests with $\alpha = 0.5$

Fig. 4. Resources usage percentage when varying the number of VNFs communicating with IoT/end-users

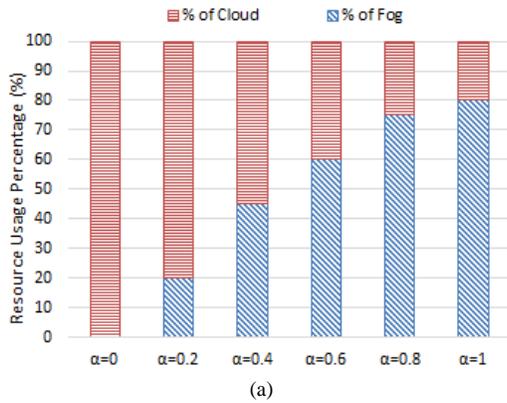
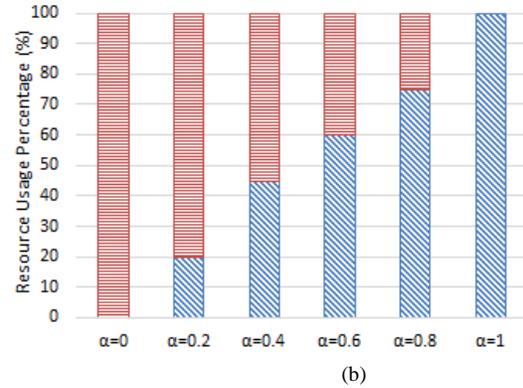

(a)                                              (b)

Fig. 3. Resources usage percentage when varying $\alpha$ considering 50 VNF-FG requests for (a) 10 nodes and (b) 20 nodes

Fig. 3 shows the percentages of the cloud/fog resources usage in TSCP for various values of α. As can be observed, when α increases, more components are placed in the fog to reduce the requests' makespan. In particular, in the infrastructure with 10 nodes, i.e., Fig. 3(a), when α = 1, some resources are still used in the cloud due to the limited number of fog nodes or due to the fog nodes' capacity limitations (from the aspect of VNF processing and communication). On the other hand, with infrastructure with 20 nodes and thus more available fog nodes (i.e., Fig. 3(b)), all the components are deployed in the fog. As α decreases, cloud resources are used more. In the extreme case, when α = 0, all components are deployed in the cloud to minimize the cost.

Fig. 4 shows the resources usage percentages in TSCP for various amounts of communication with IoT/end-users. We have changed the number of the VNFs that communicate with IoT/end-users in each VNF-FG

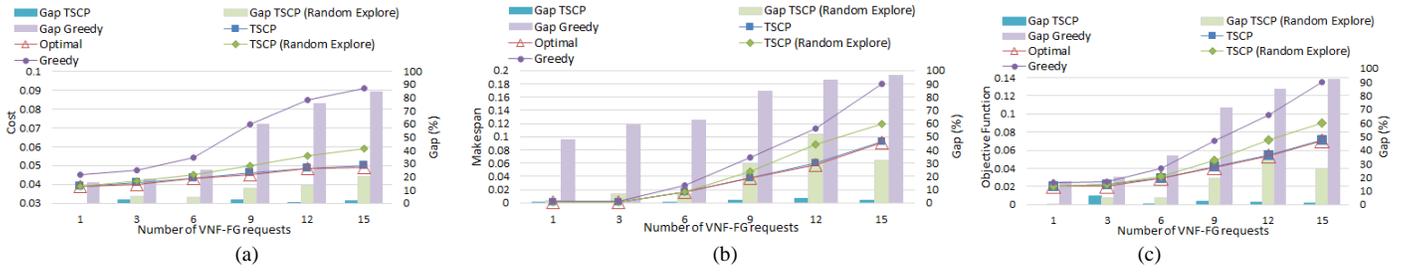

Fig. 5. Total cost (a), makespan, (b) and aggregated weighted function of cost and makespan (c) for Optimal, TSCP, Greedy, and TSCP (Random Explore), together with the gap from optimality for TSCP, TSCP (Random Explore), and Greedy for 10 nodes and up to 15 VNF-FG requests with α = 0.5

request for the case of infrastructure with 20 nodes and α = 0.5. As visible in Fig. 4, when the communications with IoT/end-users increase, more fog resources are used to reduce the communication time with IoT/end-users, and accordingly, to reduce the aggregated makespan and cost.

Fig. 5 and Fig. 6 show the cost, makespan, and their aggregated weighted function with α = 0.5 for two different scales. Fig. 5 indicates the results for infrastructure with 10 nodes and up to 15 requests. In this figure (i.e., Fig. 5) the average gap between the TSCP, Greedy, and the TSCP (Random Explore) algorithms with respect to the Optimal results is also demonstrated. As can be seen, the TSCP has cost, makespan, and objective functions that are very close to those of the Optimal result. Greedy shows the worst performance, as it selects the first available cloud/fog node without taking into account the time/cost of VNF execution. In fact, the outperformance of TSCP over Greedy is independent of the initial solution. This is because Greedy advocates a random first-fit placement without considering the objective function. On the other hand, though TSCP places the VNFs randomly in the first step, it improves the placement by applying various moves and evaluating the objective function, i.e., Eq. (16), through iterations. The TSCP outperforms TSCP (Random Explore), which demonstrates the effectiveness of the VNF execution time, hosting cost, and communication time/cost consideration in the TSCP exploration phase, as performed by the moves introduced in Section 5. Please note that the actual values for the makespan in Fig. 5(b) for points 1 and 3 are 48msec and 123msec, respectively. However, because of the normalization, in this figure, the values are close to zero.

Fig. 6 illustrates similar results for the larger scale, i.e., infrastructure with 20 nodes and up to 50 VNF-FG requests. Note that we could not get the optimal results at this scale due to its very long run time. The better

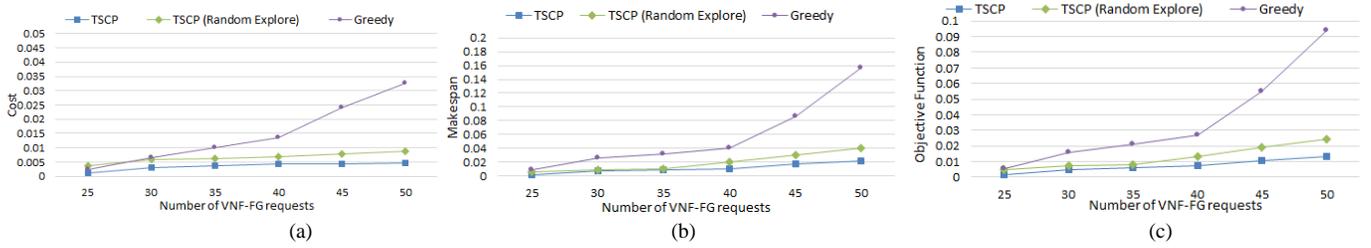
Fig. 6. Total cost (a), makespan, (b) and aggregated weighted function of cost and makespan (c) for Optimal, TSCP, Greedy, and TSCP (Random Explore) for 20 nodes and 50 VNF-FG requests with α = 0.5

performance of TCSP compared to that of the TSCP (Random Explore) and Greedy is much more remarkable here than in the smaller-scale experiment with a smaller solution space size. While the TSCP outperforms the other methods in aggregated makespan and cost by up to 47.23% (see Fig. 5), this value is up to 85% for larger-scale experiments. As can be observed, Greedy has the worst performance, since it does not consider the time/cost of VNF execution and communication when selecting the cloud/fog nodes. Similar to Fig. 5, the actual makespan values for points 25 and 30 in Fig. 6(b) are 1985.2msec and 2143.5msec, respectively. These values in Fig. 6(b) are close to zero because of the normalization. It should be noted that the actual values for the cost and makespan considering 1 to 50 requests are in the range of 300 to 2400 units of currency considering $ as a unit, and from 48msec to 3230msec, respectively.

Fig. 7 illustrates the results for the infrastructure with 10 nodes and 15 VNF-FG requests. Different types of infrastructures are considered: when the infrastructure is provided as a cloud, when it is provided as a fog, and the hybrid case consisting of both cloud and fog. As can be observed in Fig. 7(a), in every method, the cost is minimized by using only cloud, as the resources in the cloud are cheaper than those in the fog. On the other hand, the makespan is minimized by using only fog (i.e., Fig. 7(b)). This is because fog provides lower communication time than the cloud due to its proximity to IoT/end-users; a situation which leads to makespan reduction. We can also see that the best results for the aggregated weighted function of makespan and cost (i.e., Fig. 7(c)) are obtained when the components are placed on hybrid cloud/fog system for all the algorithms.

Table IV shows the computational complexity of the TSCP in comparison with the Optimal solution. The complexity values have been observed during the simulation. The TSCP clearly has a much shorter execution time than the Optimal. For the infrastructure with 20 nodes and 5 requests, the execution time of the Optimal solution exceeds 1 hour. The execution time increases as the scale of the infrastructure or the number of requests increases. For example, when the number of requests increases to 15, it took one full day to find the optimal placement, while the TSCP could find a near-optimal placement in less than 6 seconds.

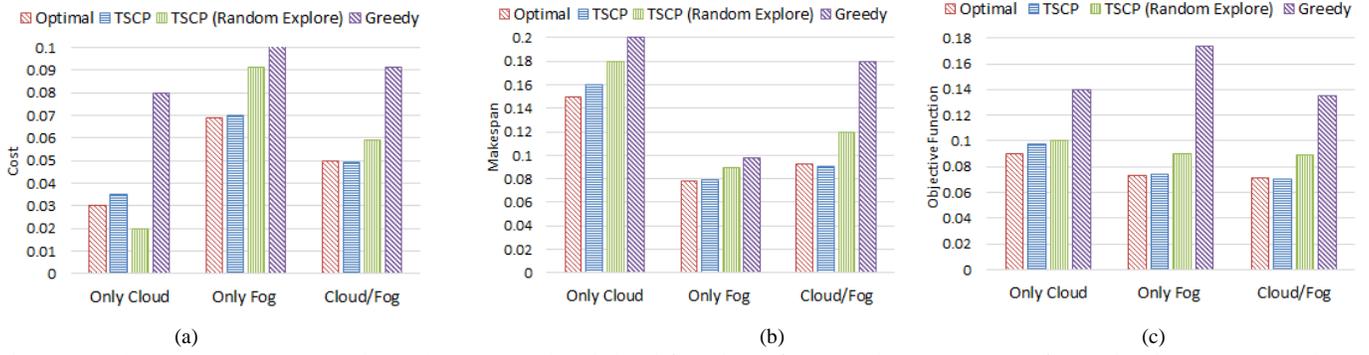

Fig. 7. Total cost (a), makespan, (b) and aggregated weighted function of cost and makespan (c) for optimal, TSCP, Greedy, and TSCP (Random Explore) for 10 nodes and 15 VNF-FG requests with α = 0.5, considering three scenarios: only cloud, only fog, cloud/fog system

Table IV. Average execution time

| Experiment Parameter | | Execution Time (sec) | |
| --- | --- | --- | --- |
| Number of Nodes | Number of VNF-FGs | *Optimal* | *Tabu Search Algorithm* |
| 10 | 5 | 4800 | 0.21 |
| 20 | 5 | 5400 | 0.58 |
| 10 | 10 | 21600 | 1.32 |
| 20 | 10 | 25560 | 2.45 |
| 10 | 15 | 54000 | 5.12 |
| 20 | 15 | > 86400 | 5.6 |
| 20 | 50 | ∞ | 57 |

## 7. CONCLUSION

This paper studies the application component placement problem in NFV-based hybrid cloud/fog systems with mobile fog nodes. The applications' components are implemented as VNFs. A structured VNF-FGs with sub-structures such as sequence, parallel, selection, and loop is established to model the execution sequence of the components. The mobility of fog nodes is modeled via RWP model. Based on the stationary analysis of RWP model, the expected execution time and cost of the components and sub-structures are calculated. These calculations are aggregated to calculate the expected application makespan and cost. The problem is modeled as ILP optimization that minimizes the aggregated makespan and cost for all requests. Tabu-based algorithm is proposed to solve the problem for large number of cloud/fog nodes and large number of requests.

The simulation results show that the proposed algorithm operates at near-optimal for small scales and improves the makespan, the cost, and the aggregated of them for larger scales. Our studies also show that the greater the communication between the application components and the IoT/end-users, the more fog resources are used to reduce the makespan. In the future, we plan to extend our work by considering that applications arrive dynamically to a system. To that end, we aim to study the transient analysis of fog nodes' mobility.

ACKNOWLEDGMENT

This work is partially funded by the CISCO CRC program (Grant #973107), the Canada Research Chair Program, and the Canadian Natural Sciences and Engineering Research Council (NSERC) through the Discovery Grant program.